# Control over Berry Curvature Dipole with Electric Field in WTe$_2$

Xing-Guo Ye,[1,*] Huiying Liu,[2,*] Peng-Fei Zhu,[1,*] Wen-Zheng Xu,[1,*] Shengyuan A. Yang,[2]
Nianze Shang,[1] Kaihui Liu,[1] and Zhi-Min Liao[1,†]

[1]*State Key Laboratory for Mesoscopic Physics and Frontiers Science Center for Nano-optoelectronics, School of Physics, Peking University, Beijing 100871, China*
[2]*Research Laboratory for Quantum Materials, Singapore University of Technology and Design, Singapore, 487372, Singapore*

Berry curvature dipole plays an important role in various nonlinear quantum phenomena. However, the maximum symmetry allowed for nonzero Berry curvature dipole in the transport plane is a single mirror line, which strongly limits its effects in materials. Here, via probing the nonlinear Hall effect, we demonstrate the generation of Berry curvature dipole by applied dc electric field in WTe$_2$, which is used to break the symmetry constraint. A linear dependence between the dipole moment of Berry curvature and the dc electric field is observed. The polarization direction of the Berry curvature is controlled by the relative orientation of the electric field and crystal axis, which can be further reversed by changing the polarity of the dc field. Our Letter provides a route to generate and control Berry curvature dipole in broad material systems and to facilitate the development of nonlinear quantum devices.

Berry curvature is an important geometrical property of Bloch bands, which can lead to a transverse velocity of Bloch electrons moving under an external electric field [1–6]. Hence, it is often regarded as a kind of magnetic field in momentum space, leading to various exotic transport phenomena, such as anomalous Hall effect (AHE) [1], anomalous Nernst effect [7], and extra phase shift in quantum oscillations [8]. The integral of Berry curvature over the Brillouin zone for fully occupied bands gives rise to the Chern number [5], which is one of the central concepts of topological physics.

Recently, Sodemann and Fu [9] proposed that the dipole moment of Berry curvature over the occupied states, known as Berry curvature dipole (BCD), plays an important role in the second-order nonlinear AHE in time-reversal-invariant materials. For transport in the *x-y* plane which is typical in experiments, the relevant BCD components form an in-plane pseudovector with $D_\alpha = \int_k f_0(\partial_\alpha \Omega_z)$ [9], where $D_\alpha$ is the BCD component along direction $\alpha$, $\bm{k}$ is the wave vector, the integral is over the Brillouin zone and with summation over the band index, $f_0$ is the Fermi distribution (in the absence of external field), $\Omega_z$ is out-of-plane Berry curvature, and $\partial_\alpha = \partial/\partial k_\alpha$. It results in a second-harmonic Hall voltage in response to a longitudinal ac probe current, which could find useful applications in high-frequency rectifiers, wireless charging, energy harvesting, and infrared detection, etc. BCD and its associated nonlinear AHE have been predicted in several material systems [9–11] and experimentally detected in systems such as two-dimensional (2D) monolayer or few-layer WTe$_2$ [12–15], Weyl semimetal TaIrTe$_4$ [16], 2D MoS$_2$, and WSe$_2$ [17–20], corrugated bilayer graphene [21], and a few topological materials [22–25]. However, a severe limitation is that BCD obeys a rather stringent symmetry constraint. In the transport plane, the maximum symmetry allowed for $D_\alpha$ is a single mirror line [9]. In several previous Letters [17–21], one needs to perform additional material engineering such as lattice strain or interlayer twisting to generate a sizable BCD. This constraint limits the available material platforms with nonzero BCD, unfavorable for the in-depth exploration of BCD-related physics and practical applications.

Recent works suggested an alternative route to obtain nonzero BCD, that is, utilizing the Berry connection polarizability to achieve a field-induced BCD, where the additional lattice engineering is unnecessary [26,27]. The Berry connection polarizability is also a band geometric quantity, related to the field-induced positional shift of Bloch electrons [28]. It is a second-rank tensor, defined as $G_{ab}(\bm{k}) = [\partial A_a^{(1)}(\bm{k})/\partial E_b]$, where $\bm{A}^{(1)}$ is the field-induced Berry connection, $\bm{E}$ is the applied electric field [28], and the superscript "(1)" represents that the physical quantity is the first order term of electric field. Then, the $E$ field induced Berry curvature is given by $\bm{\Omega}^{(1)} = \nabla_\bm{k} \times (\overleftrightarrow{\bm{G}}\bm{E})$ [27], where the double arrow indicates a second-rank tensor. This field-induced Berry curvature will lead to a field-induced BCD $D_\alpha^{(1)}$. Considering transport in the *x-y* plane and applied dc $E$ field also in the plane, we have $D_\alpha^{(1)} = \int_k f_0(\partial_\alpha \Omega_z^{(1)}) = \varepsilon_{z\gamma\mu} \int_k f_0[\partial_\alpha(\partial_\gamma G_{\mu\nu})]E_\nu$, where $\alpha, \gamma, \mu, \nu = x, y$, and $\varepsilon_{z\gamma\mu}$ is the Levi-Civita symbol. In systems where the original BCD is forbidden by the crystal symmetry, the field-induced BCD by an external $E$ field could generally be nonzero and become the dominant contribution. In such a case, the symmetry is lowered by the applied $E$ field, and the induced BCD should be linear with $E$ and its direction also controllable by the $E$ field. So far, this BCD caused by Berry connection polarizability and its field control have not been experimentally demonstrated yet, and the nonlinear Hall effect derived from this mechanism has not been observed.

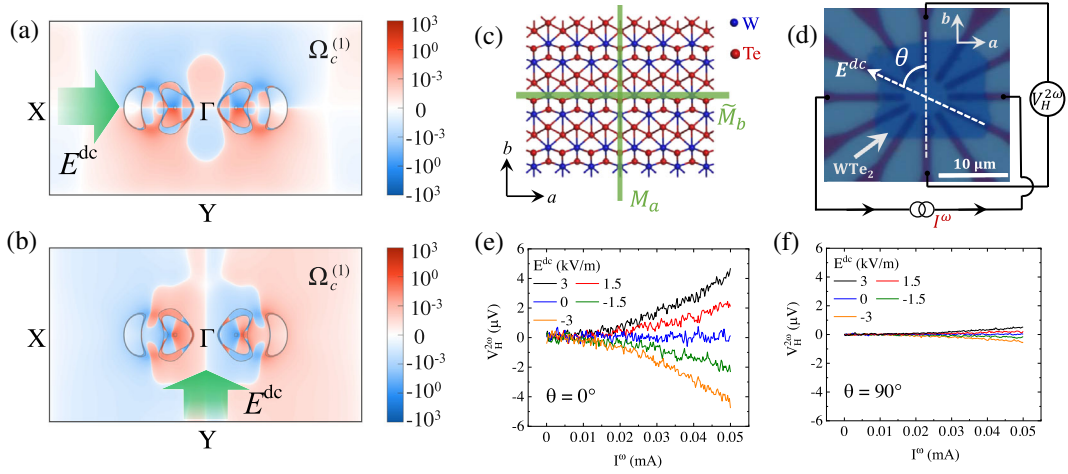

FIG. 1. (a) and (b) The field-induced Berry curvature $\Omega_c^{(1)}(\mathbf{k})$ in the $k_z = 0$ plane by a dc electric field $E^{dc} = 3$ kV/m applied along (a) $a$ or (b) $b$ axis, respectively. The unit of $\Omega_c^{(1)}(\mathbf{k})$ is Å$^2$. The green arrows indicate the direction of $\mathbf{E}^{dc}$. The gray lines depict the Fermi surface. (c) The $a$-$b$ plane of monolayer $T_d$-WTe$_2$. (d) The optical image of device S1, where an angle $\theta$ is defined. (e) and (f) The second-harmonic Hall voltage $V_H^{2\omega}$ as $\mathbf{E}^{dc}$ (e) along $b$ axis ($\theta = 0°$), and (f) along $-a$ axis ($\theta = 90°$) at 5 K. The $\mathbf{E}^\omega$ is applied along $-a$ axis, as schematized in (d).

In this Letter, we report the manipulation of electric field induced BCD due to the Berry connection polarizability. Utilizing a dc electric field $\mathbf{E}^{dc}$ to produce BCD in bulk WTe$_2$ (for which the inherent BCD is symmetry forbidden), the second-harmonic Hall voltage $V_H^{2\omega}$ is measured as a response to an applied ac current $I^\omega$. Both orientation and magnitude of the induced BCD are highly tunable by the applied $\mathbf{E}^{dc}$. Our Letter provides a general route to extend BCD to abundant material platforms with high tunability, promising for practical applications.

The WTe$_2$ devices were fabricated with circular disc electrodes (device S1) or Hall-bar shaped electrodes (device S2). The WTe$_2$ flakes were exfoliated from bulk crystal and then transferred onto the prefabricated electrodes (Supplemental Material, Note 1 [29]). The WTe$_2$ thickness of device S1 is 8.4 nm (Supplemental Material, Fig. S1 [29]), corresponding to a 12-layer WTe$_2$, and we present the results from device S1 in the main text. The crystal orientations of WTe$_2$ devices were identified by their long, straight edges [12] and further confirmed by both polarized Raman spectroscopy (Supplemental Material, Note 2 [29]) and angle-dependent transport measurements (Supplemental Material, Note 3 [29]). The electron mobility of device S1 is $\sim 4974$ cm$^2$/V s at 5 K (Supplemental Material, Note 4 [29]).

In our experiments, we use thick $T_d$-WTe$_2$ samples (thickness $\sim$8.4 nm), which have an effective inversion symmetry in the $x$-$y$ plane (which is the transport plane). This is formed by the combination of the mirror symmetry $M_a$ and the glide mirror symmetry $\tilde{M}_b$, as indicated in Fig. 1(c). The in-plane inversion leads to the absence of inherent in-plane BCD and hence the nonlinear Hall effect in bulk (see Supplemental Material, Note 5 [29] for detailed symmetry analysis). Because $\tilde{M}_b$ involves a half-cell translation along the $c$ axis and hence is broken on the sample surface, a small but nonzero intrinsic BCD may exist on the surface. In fact, such BCD due to surface symmetry breaking has already been reported [13], and is also observed in our samples, although the signal is much weaker in thicker samples (see Supplemental Material, Fig. S9 [29]).

To induce BCD in bulk WTe$_2$ through Berry connection polarizability, a dc electric field $\mathbf{E}^{dc}$ is applied in the $x$-$y$ plane. As shown in Figs. 1(a) and 1(b), the field-induced Berry curvature shows a dipolelike distribution with nonzero BCD (theoretical calculations; see Supplemental Material, Note 6 [29]). The induced BCD can be controlled by the dc $E$ field and should satisfy the following symmetry requirements. Because the presence of a mirror symmetry would force the BCD to be perpendicular to the mirror plane [9], the induced BCD $\mathbf{D}^{(1)}$ must be perpendicular to $\mathbf{E}^{dc}$ when $\mathbf{E}^{dc}$ is along the $a$ or $b$ axis. Control experiments were carried out in device S1 to confirm the above expectations. The measurement configuration is shown in Fig. 1(d) (see Supplemental Material, Fig. S2 [29], for circuit schematic). The probe ac current with ac field $\mathbf{E}^\omega$ and frequency $\omega$ was applied approximately along the $-a$ axis, satisfying $E^\omega \ll E^{dc}$, and the second-harmonic Hall voltage $V_H^{2\omega}$ was measured to reveal the nonlinear Hall effect. The $\mathbf{E}^{dc}$ that is used to produce BCD was applied along the direction characterized by the angle $\theta$, which is the angle between the direction of $\mathbf{E}^{dc}$ and the baseline of a pair of electrodes [white line in Fig. 1(d)] that is approximately along the $b$ axis. Then $\mathbf{E}^{dc}$ along $\theta = 0°$ ($b$ axis) and $\theta = 90°$ ($-a$ axis) correspond to the induced $\mathbf{D}^{(1)}$ along the $a$ axis and $b$ axis, respectively. Because the nonlinear Hall voltage $V_H^{2\omega}$ is proportional to $\mathbf{D}^{(1)} \cdot \mathbf{E}^\omega$ [9], the nonlinear Hall effect should be observed for $\mathbf{E}^\omega \| \mathbf{D}^{(1)}$ and be vanishing for $\mathbf{E}^\omega \perp \mathbf{D}^{(1)}$.

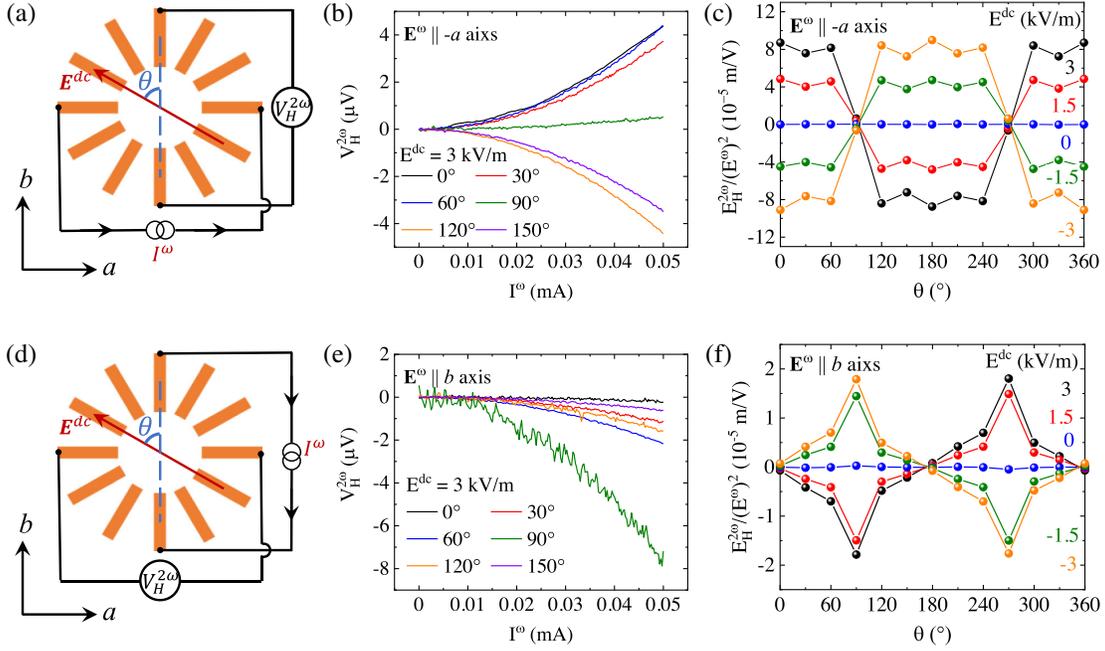

FIG. 2. (a) and (d) Measurement configuration for the second-order AHE with (a) $\mathbf{E}^\omega \| -a$ axis and (d) $\mathbf{E}^\omega \| b$ axis, respectively. The $\mathbf{E}^{dc}$, satisfying $E^{dc} \gg E^\omega$, is rotated to along various directions. (b) and (e) The second-order Hall voltage $V_H^{2\omega}$ as a function of $I^\omega$ at fixed $E^{dc} = 3$ kV/m but along various directions and at 5 K with (b) $\mathbf{E}^\omega \| -a$ axis and (e) $\mathbf{E}^\omega \| b$ axis, respectively. (c) and (f) The second-order Hall signal $[E_H^{2\omega}/(E^\omega)^2]$ as a function of $\theta$ at 5 K with (c) $\mathbf{E}^\omega \| -a$ axis and (f) $\mathbf{E}^\omega \| b$ axis, respectively.

As shown in Fig. 1(e), when $\mathbf{E}^{dc}$ along $\theta = 0°$, nonlinear Hall voltage $V_H^{2\omega}$ is indeed observed as expected. The $\mathbf{E}^{dc}$ along the $b$ axis induces BCD along the $a$ axis, leading to nonzero $V_H^{2\omega}$ since $\mathbf{E}^\omega$ is applied along the $-a$ axis. The second-order nature is verified by both the second-harmonic signal and parabolic $I$-$V$ characteristics. It is found that the nonlinear Hall voltage is highly tunable by the magnitude of $E^{dc}$. The sign reverses when $E^{dc}$ is reversed. Moreover, the nonlinear Hall voltage is linearly proportional to $E^{dc}$ (Supplemental Material [29] Fig. S11), as we expected. As for $\mathbf{E}^{dc}$ along $\theta = 90°$, as shown in Fig. 1(f), the $V_H^{2\omega}$ is much suppressed, which is at least one order of magnitude smaller than the $V_H^{2\omega}$ in Fig. 1(e). Because in this case the $\mathbf{E}^{dc}$ along the $a$ axis induces BCD along the $b$ axis, $\mathbf{E}^\omega$ is almost perpendicular to BCD, leading to negligible nonlinear Hall effect. Similar results are also reproduced in device S2 (Supplemental Material [29], Fig. S12). Such control experiments are well consistent with our theoretical expectation and confirm the validity of field-induced BCD.

Besides the crystalline axis ($\theta = 0°$ and $90°$), we also study the case when $\mathbf{E}^{dc}$ is applied along arbitrary $\theta$ directions to obtain the complete angle dependence of field-induced BCD. Here, $\mathbf{E}^\omega$ is applied along the $-a$ or $b$ axis, to detect the BCD component along the $a$ or $b$ axis, i.e., $\mathbf{D}^{(1)} = [D_a^{(1)}(\theta), D_b^{(1)}(\theta)]$, where $D_a^{(1)}$ and $D_b^{(1)}$ are the BCD components along the $a$ and $b$ axis, respectively. The measurement configurations are shown in Figs. 2(a) and 2(d). Figures 2(b) and 2(e) show the second-order Hall voltage as a function of $\theta$, with the magnitude of $\mathbf{E}^{dc}$ fixed at 3 kV/m. The second-order Hall response $[E_H^{2\omega}/(E^\omega)^2]$ is calculated by $E_H^{2\omega} = (V_H^{2\omega}/W)$ and $E^\omega = (I^\omega R_\|/L)$, where $W$ is the channel width, $R_\|$ is the longitudinal resistance, and $L$ is the channel length. As shown in Figs. 2(c) and 2(f), $[E_H^{2\omega}/(E^\omega)^2]$ demonstrates a strong anisotropy, closely related to the inherent symmetry of WTe$_2$. First of all, it is worth noting that the second-order Hall signal is negligible at $\mathbf{E}^{dc} = 0$. This is consistent with our previous analysis that the inherent bulk in-plane BCD is symmetry forbidden [26,27]. Second, $[E_H^{2\omega}/(E^\omega)^2]$ almost vanishes when $\mathbf{E}^{dc} \| \mathbf{E}^\omega$ along $a$ or $b$ axis. This is constrained by the mirror symmetries $M_a$ or $\tilde{M}_b$, forcing the BCD to be perpendicular to the mirror plane in such configurations.

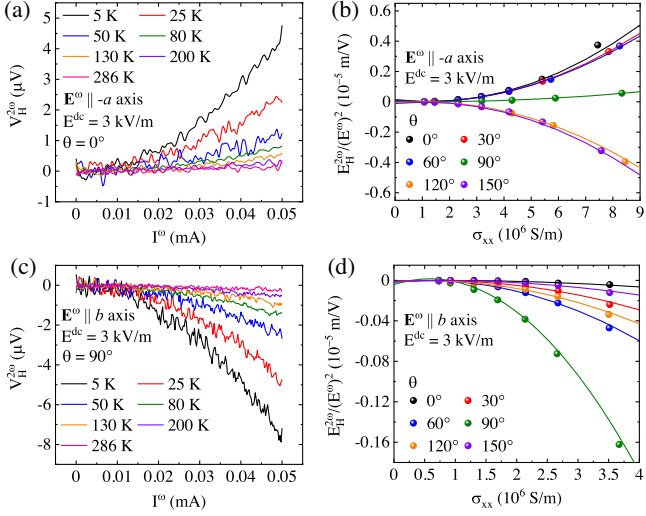

thermoelectric effect, could be safely ruled out as the main reason of the observed second-order nonlinear AHE (see Supplemental Material, Note 9 [29]). To further investigate this effect, the temperature dependence and scaling law of the second-order nonlinear Hall signal are studied. By changing the temperature, $V_H^{2\omega}$ and longitudinal conductivity $\sigma_{xx}$ were collected, where the magnitude of $\mathbf{E}^{dc}$ was fixed at 3 kV/m. Figures 3(a) and 3(c) show the $V_H^{2\omega}$ at different temperatures with $\mathbf{E}^{\omega} \| -a$ axis, $\theta=0°$ and $\mathbf{E}^{\omega}\|b$ axis, $\theta=90°$, respectively. A relatively small but nonzero second-order Hall signal is observed at 286 K. The scaling law, that is, the second-order Hall signal $[E_H^{2\omega}/(E^{\omega})^2]$ versus $\sigma_{xx}$, is presented and analyzed in Figs. 3(b) and 3(d) for different angles $\theta$. The $\sigma_{xx}$ was calculated by $\sigma_{xx}=(1/R_\|)(L/Wd)$, where $d$ is the thickness of WTe$_2$, and was varied by changing temperature. According to Ref. [42], the scaling law between $[E_H^{2\omega}/(E^{\omega})^2]$ and $\sigma_{xx}$ satisfies $[E_H^{2\omega}/(E^{\omega})^2]=C_0+C_1\sigma_{xx}+C_2\sigma_{xx}^2$. The coefficients $C_2$ and $C_1$ involve the mixing contributions from various skew scattering processes [42–45], such as impurity scattering, phonon scattering, and mixed scattering from both phonons and impurities [42]. $C_0$ is mainly contributed by the intrinsic mechanism, i.e., the field-induced BCD here. As shown in Figs. 3(b) and 3(d), the scaling law is well fitted for all angles $\theta$.

It is found that $C_0$ shows strong anisotropy (Supplemental Material [29], Fig. S18), indicating the field-induced BCD is also strongly dependent on angle $\theta$. The value of field-induced BCD can be estimated through $D=(2\hbar^2 n/m^* e)[E_H^{2\omega}/(E^\omega)^2]$ [12], where $\hbar$ is the reduced Planck constant, $e$ is the electron charge, $m^*=0.3m_e$ is the effective electron mass, $n$ is the carrier density. Here, we replace the $[E_H^{2\omega}/(E^\omega)^2]$ by the coefficient $C_0$ from the scaling law fitting. The two components of BCD along the $a$ and $b$ axes, denoted as $D_a^{(1)}$ and $D_b^{(1)}$, are calculated from the fitting curves with the magnitude of $\mathbf{E}^{dc}$ fixed at 3 kV/m under the $\mathbf{E}^{\omega}\|-a$ axis and the $\mathbf{E}^{\omega}\|b$ axis, respectively. As shown in Figs. 4(a) and 4(b), it is found that $D_a^{(1)}$ shows a $\cos\theta$ dependence on $\theta$, whereas $D_b^{(1)}$

FIG. 3. (a) and (c) The second-harmonic Hall voltage at various temperatures with the magnitude of $\mathbf{E}^{dc}$ fixed at 3 kV/m (a) under $\mathbf{E}^{\omega}\|-a$ axis, $\theta=0°$ and (c) under $\mathbf{E}^{\omega}\|b$ axis, $\theta=90°$. (b),(d) Second-order Hall signal $[E_H^{2\omega}/(E^\omega)^2]$ as a function of $\sigma_{xx}$ (b) under $\mathbf{E}^{\omega}\|-a$ axis and (d) under $\mathbf{E}^{\omega}\|b$ axis at various $\theta$ with the magnitude of $\mathbf{E}^{dc}$ fixed at 3 kV/m. The temperature range for the scaling law in (b) and (d) is 50–286 K.

Thus, when $\mathbf{E}^{dc}\|\mathbf{E}^{\omega}$ along the $a$ or $b$ axis, the induced BCD is perpendicular to $\mathbf{E}^{dc}$ and $\mathbf{E}^{\omega}$, satisfying $\mathbf{D}^{(1)}\cdot\mathbf{E}^{\omega}=0$, which leads to almost vanished second-order Hall signals. Moreover, $[E_H^{2\omega}/(E^\omega)^2]$ exhibits a sensitive dependence on the angle $\theta$, indicating the BCD is highly tunable by the orientation of $\mathbf{E}^{dc}$. A local minimum of $[E_H^{2\omega}/(E^\omega)^2]$ is found at an intermediate angle around $\theta=30°$ when $\mathbf{E}^{\omega}\|-a$ axis in Fig. 2(c). This is because $[E_H^{2\omega}/(E^\omega)^2]$ depends not only on $(\mathbf{D}^{(1)}\cdot\widehat{\mathbf{E}^\omega})$, i.e., the projection of the pseudovector $\mathbf{D}^{(1)}$ to the direction of $\mathbf{E}^\omega$, but also on the anisotropy of conductivity in WTe$_2$. The two terms show different dependence on the angle $\theta$, leading to a local minimum around $\theta=30°$.

Through control experiments and symmetry analysis, the extrinsic effects, such as diode effect, thermal effect, and shows a $\sin\theta$ dependence. Such angle dependence is well consistent with the theoretical predications (see Supplemental Material [29], Note 6). According to the two components $D_a^{(1)}$ and $D_b^{(1)}$, the field induced BCD vector of $\mathbf{D}^{(1)}$ is synthesized for $\mathbf{E}^{dc}$ along various directions, as presented in Fig. 4(c). It is found that both the magnitude and orientation of the field-induced BCD are highly tunable by the dc field.

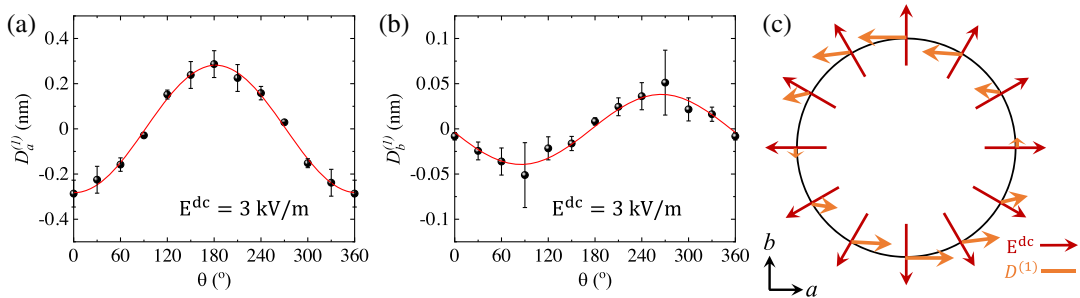

FIG. 4. The induced Berry curvature dipole as a function of $\theta$ with the magnitude of $\mathbf{E}^{dc}$ fixed at 3 kV/m for (a) the component along $a$ axis, $D_a^{(1)}$ and (b) the component along $b$ axis, $D_b^{(1)}$. (c) The relationship between the field-induced Berry curvature dipole $\mathbf{D}^{(1)}$ and the applied $\mathbf{E}^{dc}=3$ kV/m along different directions. The scale bar of $D^{(1)}$ is 0.2 nm.

In summary, we have demonstrated the generation, modulation, and detection of the induced BCD due to the Berry connection polarizability in $WTe_2$. It is found that the direction of the generated BCD is controlled by the relative orientation between the applied $\mathbf{E}^{dc}$ direction and the crystal axis, and its magnitude is proportional to the intensity of $\mathbf{E}^{dc}$. Using independent control of the two applied fields, our Letter demonstrates an efficient approach to probe the nonlinear transport tensor symmetry, which is also helpful for full characterization of nonlinear transport coefficients. Moreover, the manipulation of BCD up to room temperature by electric means without additional symmetry breaking will greatly extend the BCD-related physics [46,47] to more general materials and should be valuable for developing devices utilizing the geometric properties of Bloch electrons.

This work was supported by National Key Research and Development Program of China (No. 2018YFA0703703), National Natural Science Foundation of China (Grants No. 91964201 and No. 61825401), and Singapore MOE AcRF Tier 2 (MOE-T2EP50220-0011). We are grateful to Dr. Yanfeng Ge at SUTD for inspired discussions.

*These authors contributed equally to this work.
†liaozm@pku.edu.cn

# Supplemental Material for

# Control over Berry curvature dipole with electric field in WTe$_2$

Xing-Guo Ye[1,+], Huiying Liu[2,+], Peng-Fei Zhu[1,+], Wen-Zheng Xu[1,+], Shengyuan A. Yang[2], Nianze Shang[1], Kaihui Liu[1], and Zhi-Min Liao[1,*]

[1] State Key Laboratory for Mesoscopic Physics and Frontiers Science Center for Nano-optoelectronics, School of Physics, Peking University, Beijing 100871, China.

[2] Research Laboratory for Quantum Materials, Singapore University of Technology and Design, Singapore, 487372, Singapore.

+ These authors contributed equally.

* Email: liaozm@pku.edu.cn

**This file contains supplemental Figures S1-S18 and Notes 1-10.**

**Note 1:** Device fabrication, experimental and calculation methods.

**Note 2:** Polarized Raman spectroscopy of WTe$_2$.

**Note 3:** Angle-dependent longitudinal resistance and third-order nonlinear Hall effect.

**Note 4:** Magnetotransport properties of WTe$_2$.

**Note 5:** Symmetry analysis of WTe$_2$.

**Note 6:** Theoretical analysis and calculations of field-induced Berry curvature dipole.

**Note 7:** Electric field dependence of second-order Hall signals.

**Note 8:** Control experiments in device S2.

**Note 9:** Discussions of other possible origins of the second order AHE.

**Note 10:** Angle dependence of parameter C$_0$ obtained from the fittings of scaling law.

## Supplemental Note 1: Device fabrication, experimental and calculation methods.

### 1) Device fabrication

The WTe$_2$ flakes were exfoliated from bulk crystal by scotch tape and then transferred onto the polydimethylsiloxane (PDMS). The PDMS was then covered onto a Si substrate with 285 nm-thick SiO$_2$, where the Si substrate was precleaned by air plasma, and further heated for about 1 minute at 90°C to transfer the WTe$_2$ flakes onto Si substrate. Disk and Hall bar-shaped Ti/Au electrodes (around 10 nm thick) were prefabricated on individual SiO$_2$/Si substrates with e-beam lithography, metal deposition and lift-off. Exfoliated BN (around 20 nm thick) and WTe$_2$ flakes (around 5-20 nm thick) were sequentially picked up and then transferred onto the Ti/Au electrodes using a polymer-based dry transfer technique [30]. The atomic force microscope image of device S1 is shown in **Fig. S1**. The thickness of this sample is 8.4 nm, corresponding to a 12-layer WTe$_2$. The whole exfoliation and transfer processes were done in an argon-filled glove box with O$_2$ and H$_2$O content below 0.01 parts per million to avoid sample degeneration.

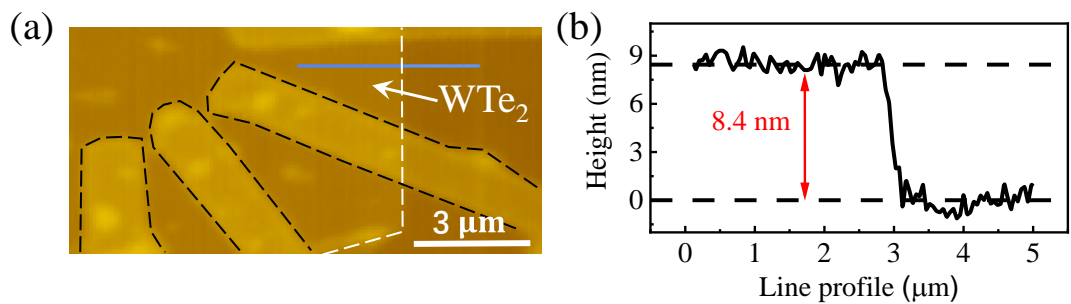

**Figure S1: (a) The atomic force microscope image of device S1. (b) The line profile shows the thickness of the WTe$_2$ sample is 8.4 nm.**

## 2) Electrical transport measurements and circuit schematic

All the transport measurements were carried out in an Oxford cryostat with a variable temperature insert and a superconducting magnet. First-, second- and third-harmonic signals were collected by standard lock-in techniques (Stanford Research Systems Model SR830) with frequency ω. Frequency ω equals 17.777 Hz unless otherwise stated.

The circuit schematic with multiple sources in experiments is depicted in **Fig. S2**. The a.c. and d.c. sources are both effective current sources. The original SR830 a.c. source is a voltage source. In experiments, we connected the SR830 voltage source and a protective resistor with resistance value $R_p$ in series ($R_p = 100$ kΩ for device S1 and $R_p = 10$ kΩ for device S2), as shown in **Fig. S2**. The resistance of WTe$_2$ channel is in the order of 10 Ω, much less than $R_p$, which makes the SR830 source an effective current source with excitation current $I^\omega \cong U^\omega / R_p$, where $U^\omega$ is the source voltage.

The Keithley 2400 current source is used for the d.c source. As shown in **Fig. S2**, the positive and negative terminals of the Keithley source are connected to a pair of diagonal electrodes to form a loop circuit, *i.e.*, a floating loop. The d.c. electric field is obtained by $E^{dc} = \frac{I^{dc} R_\theta}{L}$, where $I^{dc}$ is the applied d.c. current, $R_\theta$ is the resistance of WTe$_2$ along direction $\theta$, and $L$ is the channel length of WTe$_2$. The impedance of the floating Keithley source to ground is measured to be ~60 MΩ. While, the negative terminal of SR830 source is directly connected to the ground.

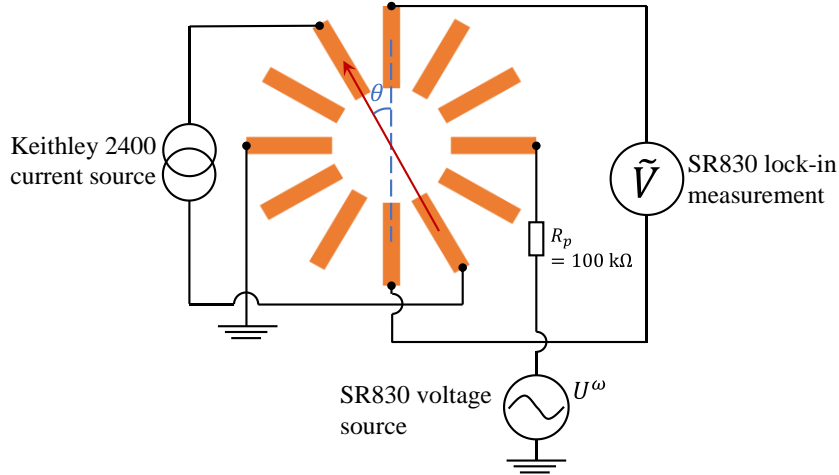

**Figure S2: Schematic structure of the circuit for measurements in device S1.**

### 3) Spectral purity of lock-in measurements

For the lock-in measurements, the used integration time is 300 ms and the filter roll-off is 24 dB/octave, that is, the cutoff (-3 dB) frequency for the low-pass filter is 0.531 Hz and the filter roll-off is 24 dB per octave. For our lock-in measurements, the narrow detection bandwidth ($\pm 0.531$ Hz) effectively avoided the spectral leakage.

The spectral purity of the lock-in homodyne circuit is verified by the control experiments of the lock-in measurements of a resistor. The first-, second- and third-harmonic voltages of a resistor with resistance ~100 Ω are measured using the same frequency (17.777 Hz), integration time (300 ms) and filter roll-off (24 dB/octave) as used in experiments, as shown in **Fig. S3**. The first-harmonic voltage shows linear dependence on the alternating current, consistent with the resistance value ~100 Ω. The second- and third-harmonic voltages are four orders of magnitude smaller than the first-harmonic voltage, which indicates the high purity of spectrum of the lock-in homodyne circuit.

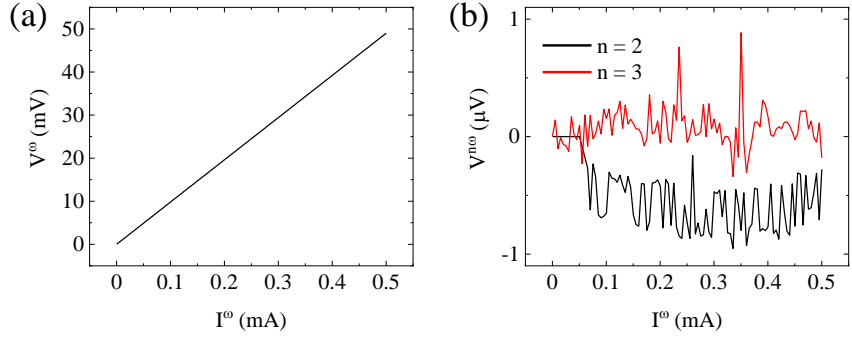

**Figure S3: Lock-in measurements for a resistor with resistance ~100 Ω.**

**a,** The first-harmonic voltage versus the alternating current.

**b,** The second- and third-harmonic voltages versus the alternating current.

**4) Validity of electrical measurements with the two sources**

In our experiments, the Keithley source is used as the d.c. current source, which has an output impedance ~20 MΩ. The a.c. current source is realized by connecting a resistor $R_p$ in series ($R_p = 100$ kΩ for device S1 and $R_p = 10$ kΩ for device S2) in series with the SR830 voltage source. Both the a.c. and d.c. current sources have effectively large output impedance comparing to the sample resistance ~10 Ω, so that they can be considered as independent current sources. These two current sources can be applied to the device simultaneously, having well-defined potential differences. To further confirm the validity of our electrical measurements with the two current sources, we design a test circuit, as shown in **Fig. S4(a)**. The a.c. current flowing through $R_2$ was calculated by measuring the first-harmonic voltage $V^\omega$ of $R_2$ and $I^\omega = V^\omega/R_2$. The d.c. current is applied by the Keithley current source and is measured by measuring the d.c. voltage $V^{dc}$ of $R_2$ and $I^{dc} = V^{dc}/R_2$. As shown in **Fig. S4(b)**, where the a.c. voltage of SR830 source is fixed at 1 V, it is found that the $I^\omega$ is unchanged when varying the d.c. current by Keithley source, while measured $I^{dc}$ is almost the same as the output current of the Keithley source. In **Fig. S4(c)**, where the d.c. current of Keithley source is fixed, it is found that $I^\omega$ well satisfies $I^\omega = U^\omega/(R_1 + R_2 + R_3) \cong U^\omega/R_p$ with $U^\omega$ as the SR830 source voltage and $R_p = R_1$. These results clearly confirm the a.c. and d.c. sources are effectively independent with negligible current shunt between each other.

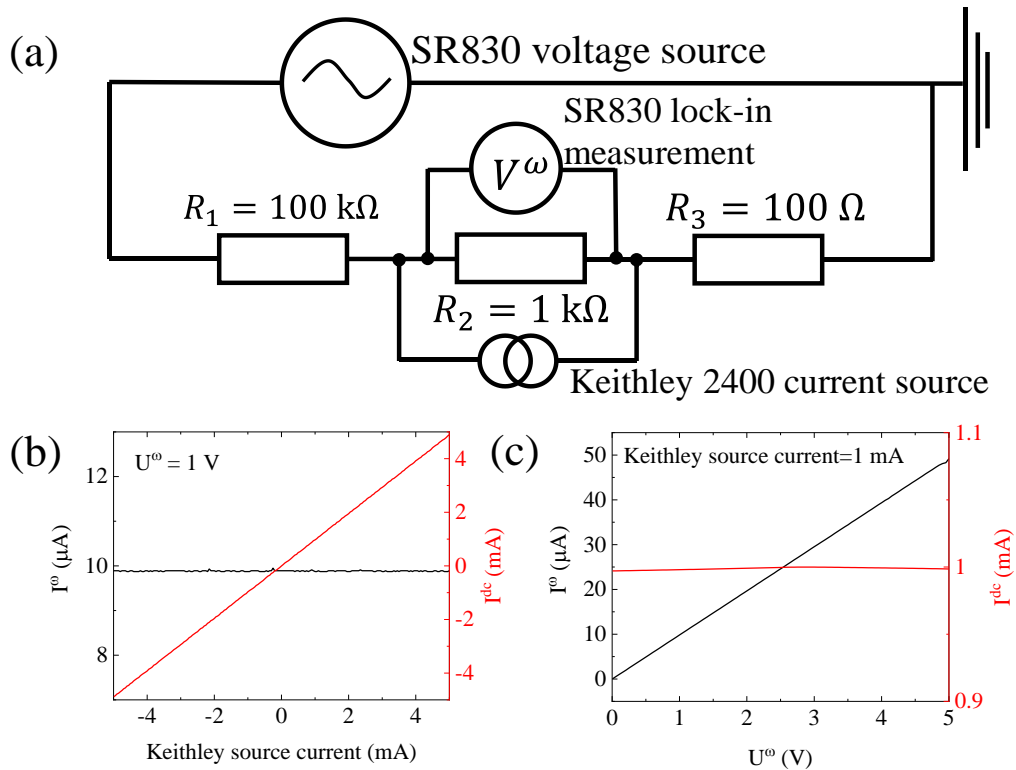

**Figure S4: Validity of the electrical measurements with two sources.**

**a,** Schematic of the test circuit.

**b,** The $I^\omega$ and $I^{dc}$ as a function of the Keithley source current with SR830 source voltage $U^\omega$ fixed at 1 V.

**c,** The $I^\omega$ and $I^{dc}$ as a function of the SR830 source voltage $U^\omega$ with Keithley source current fixed at 1 mA.

## 5) Calculation methods

First-principles calculations were performed to reveal the properties of the Berry connection polarizability tensor and field-induced Berry curvature dipole in WTe$_2$. The electronic structures were carried out in the framework of density functional theory as implemented in the Vienna ab initio simulation package [31,32] with the projector augmented wave method [33] and Perdew, Burke, and Ernzerhof exchange correlation functionals [34]. For the convergence of the results, the spin–orbit coupling was included self-consistently in the calculations of electronic structures with the kinetic energy cutoff of 600 eV and Monkhorst-Pack $k$ mesh of 14 × 8 × 4. We used $d$ orbitals of W atom and $p$ orbitals of Te atoms to construct Wannier functions [35]. While evaluating the band geometric quantities, we consider the finite temperature effect in the distribution function and a lifetime broadening of $k_B T$ with $T = 5$ K.

### Supplemental Note 2: Polarized Raman spectroscopy of WTe$_2$.

The crystalline orientation of WTe$_2$ device was determined by the polarized Raman spectroscopy in the parallel polarization configuration [36]. **Figure S5** shows the polarized Raman spectrum of device S2 as an example. The optical image of device S2 is displayed in **Fig. S5(a)**. Raman spectroscopy was measured with 514 nm excitation wavelengths through a linearly polarized solid-state laser beam. The polarization of the excitation laser was controlled by a quarter-wave plate and a polarizer. We collected the Raman scattered light with the same polarization as the excitation laser. A typical Raman spectroscopy of device S2 is shown in **Fig. S5(b)**, where five Raman peaks are identified, belonging to the A1 modes of WTe$_2$ [36]. We further measured the polarization dependence of intensities of peaks P2 and P11 [denoted in **Fig. S5(b)**] in **Figs. S5(c)** and **S5(d)**, respectively. Based on previous reports [36], the polarization direction with maximum intensity was assigned as the *b* axis. The measured crystalline orientation is further indicated in the optical image [**Fig. S5(a)**], where the applied a.c. current is approximately parallel to *a* axis.

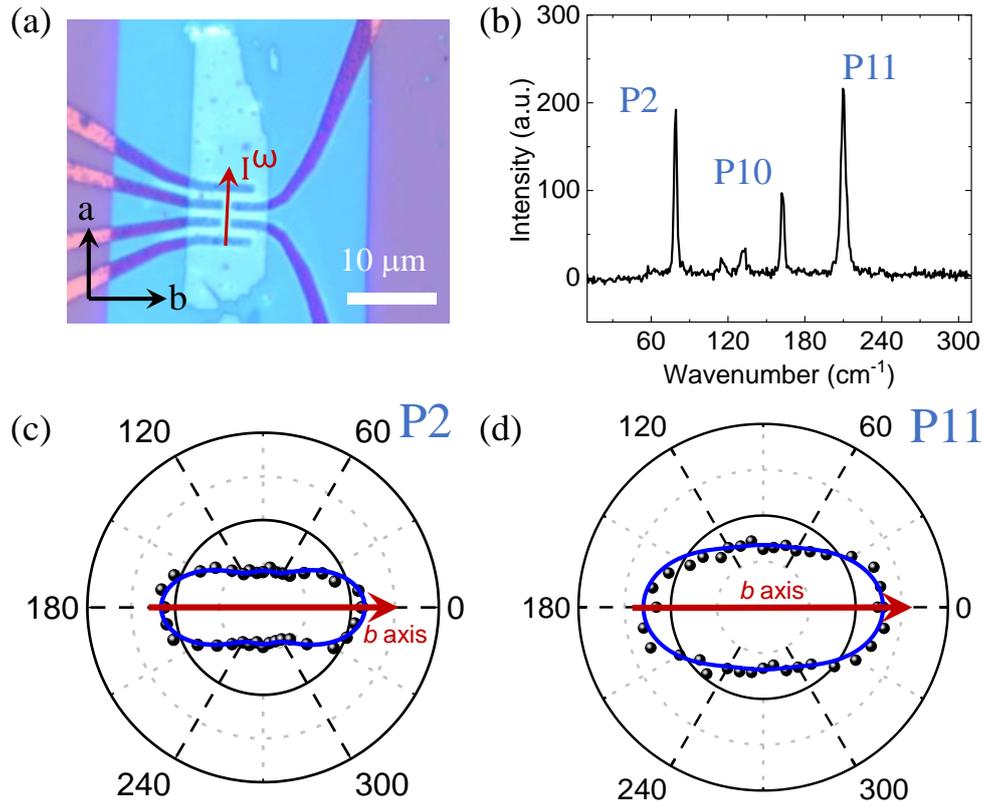

**Figure S5: Polarized Raman spectroscopy of WTe$_2$ to determine the crystalline orientation.**

**a,** Optical image of device S2. The crystalline axes, *i.e.*, *a* axis and *b* axis, determined by the polarized Raman spectroscopy, are denoted by the black arrows. The applied a.c. current is also noted by the red arrow, which is approximately aligned with *a* axis.

**b,** A typical Raman spectrum measured with 514 nm excitation wavelengths, where the polarization direction is approximately along *b* axis. Five Raman peaks are observed, which belong to the A1 modes of WTe$_2$ [36].

**c,d,** Polarization dependence of intensities of peaks (**c**) P2 and (**d**) P11. Here the polarization angle takes 0° along the *b* axis, along which maximum intensity is observed [36].

## Supplemental Note 3: Angle-dependent longitudinal resistance and third-order nonlinear Hall effect.

The third-order anomalous Hall effect (AHE) is investigated in device S1, as shown in **Fig. S6(a)**. By exploiting the circular disc electrode structure, the angle-dependence of the third-order AHE is measured. It shows highly sensitive to the crystalline orientation, as shown in **Fig. S6(c)**, which inherits from the intrinsic anisotropy of WTe$_2$ [26]. Based on the symmetry of WTe$_2$ [26], the third-order AHE shows angle-dependence following the formula

$$\frac{E_H^{3\omega}}{(E^\omega)^3} \propto \frac{\cos(\theta-\theta_0)\sin(\theta-\theta_0)[(\chi_{22}r^4-3\chi_{12}r^2)\sin^2(\theta-\theta_0)+(3\chi_{21}r^2-\chi_{11})\cos^2(\theta-\theta_0)]}{(\cos^2(\theta-\theta_0)+r\sin^2(\theta-\theta_0))^3},$$

where $E_H^{3\omega} = \frac{V_H^{3\omega}}{W}$, $E^\omega = \frac{I^\omega R_\parallel}{L}$, $V_H^{3\omega}$ is the third-harmonic Hall voltage, $I^\omega$ is the applied a.c. current, $R_\parallel$ is the longitudinal resistance, $W$ and $L$ are channel width and length, respectively, $r$ is the resistance anisotropy, $\chi_{ij}$ are elements of the third-order susceptibility tensor, $\theta_0$ is the angle misalignment between $\theta = 0°$ and crystalline $b$ axis. The fitting curve for this angle dependence is shown by the red line in **Fig. S6(c)**, which yields the misalignment $\theta_0$ ~1.5°. In addition to the third-order AHE, the longitudinal ($R_\parallel$) resistance also shows strong anisotropy [13], as shown in **Fig. S6(b)**, following

$$R_\parallel(\theta) = R_b \cos^2(\theta - \theta_0) + R_a \sin^2(\theta - \theta_0),$$

consistent with previous results [13], where $R_a$ and $R_b$ are resistance along crystalline $a$ and $b$ axis, respectively.

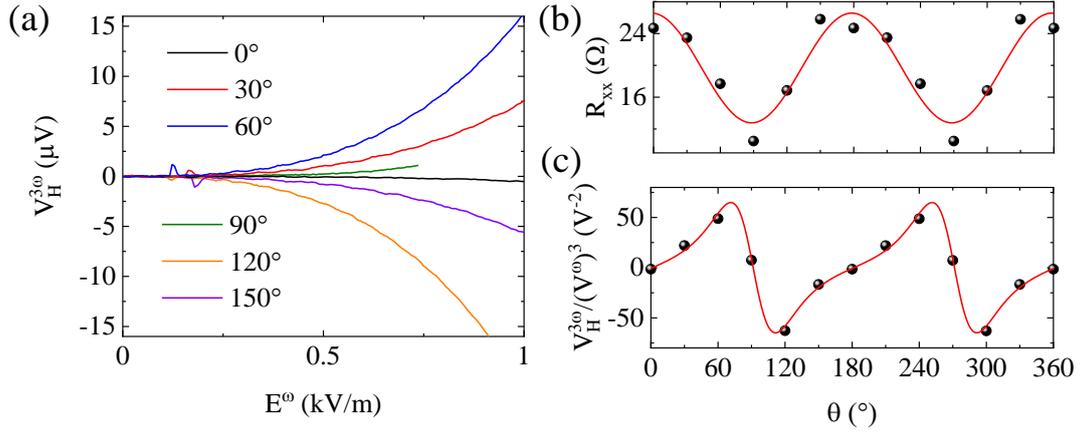

**Figure S6: Angle-dependence of third-order nonlinear Hall effect in device S1 at 5 K.**

**a,** The third-harmonic anomalous Hall voltages at various $\theta$. Here $\theta$ is defined as the relative angle between the alternating current and the baseline (approximately along *b* axis).

**b,c,** (**b**) $R_{xx}$ and (**c**) third-order Hall signal $\frac{E_H^{3\omega}}{(E^\omega)^3}$ as a function of $\theta$, respectively.

**Supplemental Note 4: Magnetotransport properties of WTe₂.**

The magneto-transport properties of the device S1 were investigated. **Figure S7(a)** shows the resistivity as a function of temperature. The resistivity decreases upon decreasing temperature with a residual-resistivity at low temperatures, showing typical metallic behaviors. **Figure S7(b)** shows the magnetoresistance (MR) and Hall resistance as a function of magnetic field. MR is defined as $\frac{R_{xx}(B)-R_{xx}(0)}{R_{xx}(0)} \times 100\%$. The low residual resistance and large, non-saturated MR indicate the high quality of the WTe₂ devices [37,38]. The carrier mobility of device S1 is estimated as high as $4974.4 \text{ cm}^2/(\text{V} \cdot \text{s})$. Moreover, resistance oscillations due to the formations of Landau levels are also observed, as shown in **Fig. S7(c)**, indicative of the high crystal quality. The oscillation $\Delta R_{xx}$ is obtained by subtracting a parabolic background. The fast Fourier transform (FFT) is performed, as shown in **Fig. S7(d)**. Three frequencies are observed, indicating the multiple Fermi pockets in WTe₂, which is consistent with previous work [37-39]. The dominant peak of FFT $f_1$ is around 44 T.

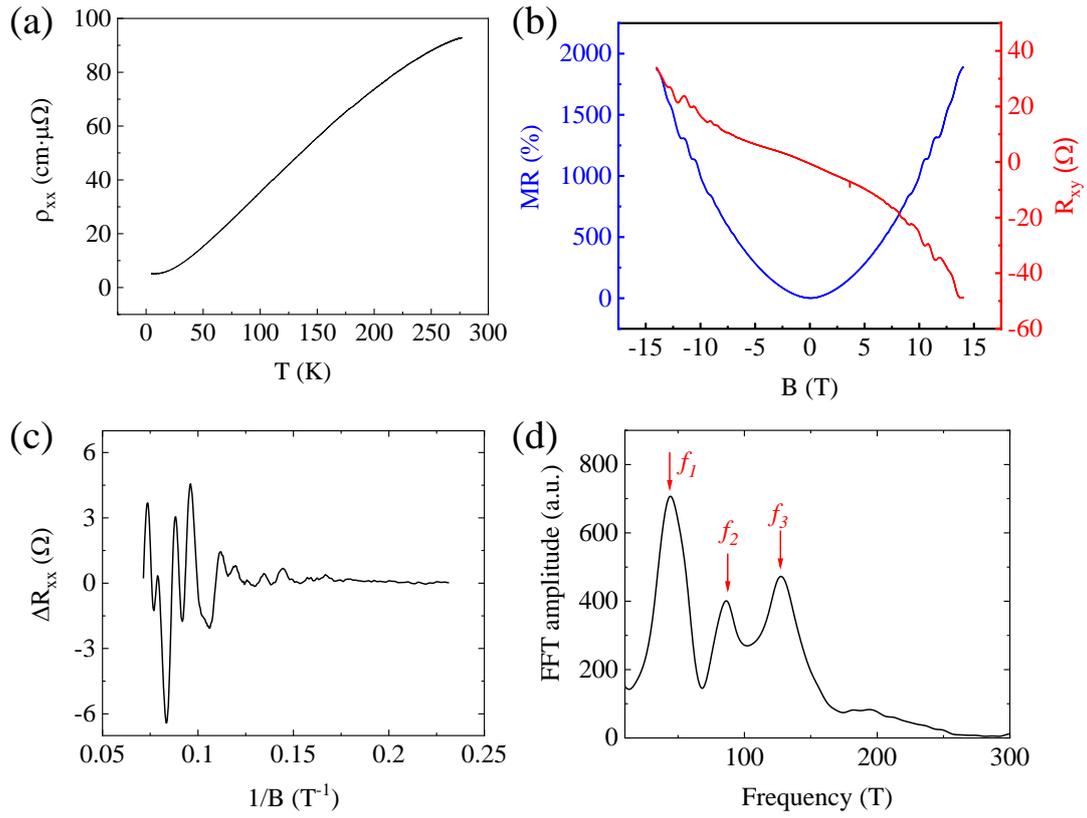

**Figure S7: Transport properties of the device S1.**

**a,** The resistivity as a function of temperature.

**b,** Magnetoresistance and Hall resistance at 5 K.

**c,** Oscillations of R$_{xx}$ at 5 K. The $\Delta R_{xx}$ is obtained by subtracting a parabolic background.

**d,** The FFT analysis of $\Delta R_{xx}$ oscillations, where three peaks are obtained.

**Supplemental Note 5: Symmetry analysis of WTe$_2$.**

T$_d$-WTe$_2$ has a distorted crystal structure with low symmetry. Here we analyze the thickness dependence of the symmetry in WTe$_2$ in details. **Figure S8(a)** shows the *b-c* plane of monolayer WTe$_2$. Each monolayer consists of a layer of W atoms sandwiched between two layers of Te atoms, denoted as Te1 (denoted in yellow) and Te2 (denoted in red), respectively. The inversion symmetry of the monolayer is approximately satisfied, and Te1 is equivalent to Te2. The presence of inversion symmetry forces Berry curvature dipole (BCD) to be zero. However, as a perpendicular displacement field is applied to break the inversion symmetry, the Te1 is no longer equivalent to Te2. As shown in the bottom of **Fig. S8(a)**, an in-plane electric polarization along *b* axis can be induced by the out-of-plane displacement field. The electric polarization along *b* axis plays a similar role as the d.c. electric field in our work, leading to nonzero BCD along *a* axis.

Nonzero BCD in bilayer WTe$_2$ origins from crystal symmetry breaking. The largest symmetry in bilayer WTe$_2$ is a single mirror symmetry $M_a$ with *bc* plane as mirror plane. As shown in **Fig. S8(b)**, the stacking between the two layers makes bilayer WTe$_2$ inversion symmetry breaking. Under inversion operation, the top and bottom layers are swapped, which fails to coincide with each other. As shown in **Fig. S8(b)**, Te1 is not equivalent to Te2 due to the stacking arrangement in bilayer. Therefore, an in-plane electric polarization **P** along *b* axis exists, similar to the case in monolayer with an out-of-plane displacement field. The polarization **P** is able to induce nonzero BCD along the perpendicular crystalline axis, *i.e.*, along *a* axis.

In fact, such in-plane polarization **P** along *b* axis in monolayer and bilayer WTe$_2$ is already evidenced by the circular photogalvanic effect [14]. The symmetry breaking induced polarization is also confirmed in various 2D materials, such as WSe$_2$/black phosphorus heterostructures [40].

In trilayer and thicker WTe$_2$, as shown in **Fig. S8(c)**, the Te1 and Te2 are equivalent in bulk, leading to vanished electric polarization. The in-plane inversion symmetry in bulk forbids the presence of in-plane BCD. However, the inversion is broken on surface. Therefore, for trilayer and thicker WTe$_2$, a small but nonzero BCD may occur on surface.

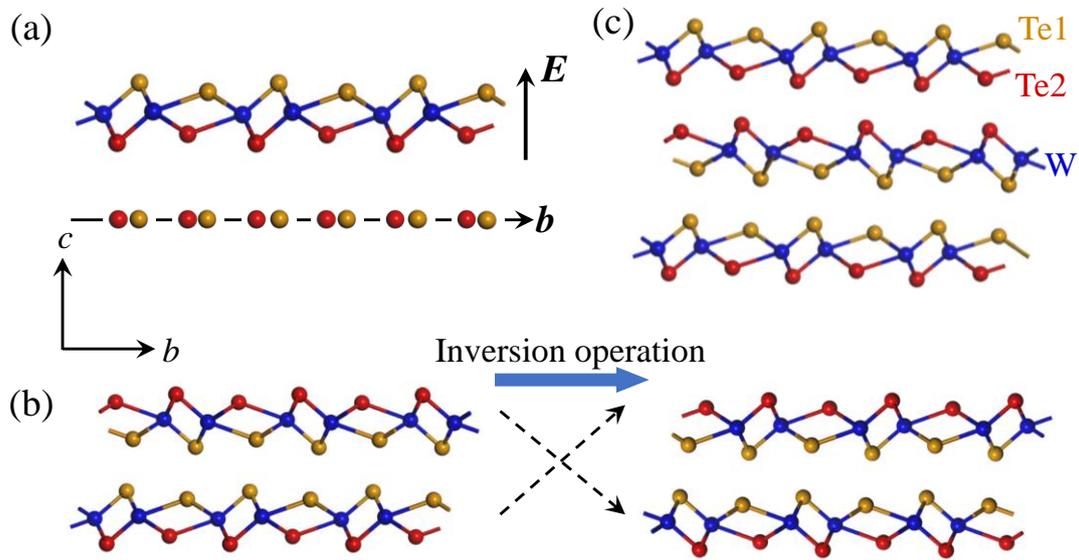

**Figure S8: Crystal structure of T$_d$-WTe$_2$.**

**a,** *b-c* plane of monolayer T$_d$-WTe$_2$.

**b,** *b-c* plane of bilayer T$_d$-WTe$_2$. The stacking arrangement breaks the inversion symmetry.

**c,** *b-c* plane of trilayer T$_d$-WTe$_2$.

Importantly, the surface BCD and it induced second-order AHE in few-layer WTe$_2$

is reported in Ref. [13], which is also observed in our device. We measured the second-order AHE without the application of $E^{dc}$ in a WTe$_2$ device, as shown in **Fig. S9**. This second-order AHE is observable when applying $I^\omega$ in the order of 1 mA. By comparison, the second-order AHE induced by d.c. field is observable when applying $I^\omega$ smaller than 0.05 mA (Fig. 1 of main text). The calculated BCD along *a* axis $D_a$ without the application of $E^{dc}$ is ~0.03 nm, which is one order of magnitude smaller than $D_a^{(1)}$ ~0.29 nm measured under $E^{dc}$ = 3kV/m (Fig. 4 of main text). These results confirm the validity of $E^{dc}$ induced BCD in our work.

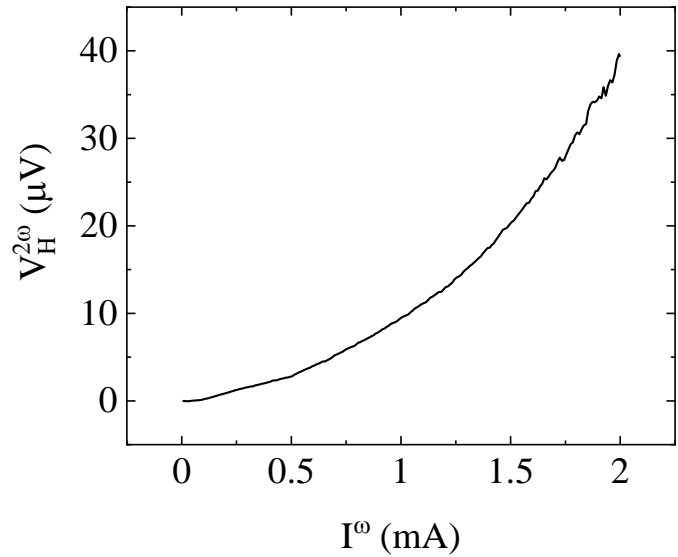

**Figure S9: The second-order AHE without external d.c. electric field in WTe$_2$ at 1.8 K.**

## Supplemental Note 6: Theoretical analysis and calculations of field-induced Berry curvature dipole.

The electric field-induced Berry curvature depends on the Berry connection polarizability tensor and the applied d.c. field with the relation that

$$\mathbf{\Omega}^{(1)} = \mathbf{\nabla}_{\mathbf{k}} \times (\overleftrightarrow{\mathbf{G}} \mathbf{E}^{dc}),$$

$$\Omega_\beta^{(1)}(n, \mathbf{k}) = \varepsilon_{\beta\gamma\mu}[\partial_\gamma G_{\mu\nu}(n, \mathbf{k})]E_\nu^{dc},$$

with $G_{\mu\nu}(n, \mathbf{k}) = 2e\mathrm{Re}\sum_{m\neq n}\frac{(A_\mu)_{nm}(A_\nu)_{mn}}{\varepsilon_n - \varepsilon_m}$, where $A_{mn}$ is the interband Berry connection and $e$ is the electron charge. The superscript "(1)" represents that the physical quantity is the first order term of electric field. Here the Greek letters refer to the spatial directions, $m, n$ refer to the energy band indices, $\varepsilon_{\beta\gamma\mu}$ is the Levi-Civita symbol, and $\partial_\gamma$ is short for $\partial/\partial k_\gamma$. The Berry connection polarizability tensor of WTe$_2$ is calculated and shown in **Figs. S10(a)-(c)**. From the definition, the field-induced BCD is

$$D_{\alpha\beta}^{(1)} = \int_k [d\mathbf{k}] f_0 \left(\partial_\alpha \Omega_\beta^{(1)}\right) = \varepsilon_{\beta\gamma\mu} \int_k [d\mathbf{k}] f_0 [\partial_\alpha(\partial_\gamma G_{\mu\nu})]E_\nu^{dc},$$

where $\int_k [d\mathbf{k}] = \sum_n \frac{1}{(2\pi)^3} \iiint d\mathbf{k}$ is taken over the first Brillouin zone of the system and summed over all energy bands.

In two-dimensional systems, $\mathbf{\Omega}^{(1)}$ is constrained to the out of plane direction, and BCD behaves as a pseudo vector in the plane. Here we choose our coordinate frame along the crystal principal axes $a, b, c$. By applying a d.c. electric field $\mathbf{E}^{dc} = (E_a^{dc}, E_b^{dc})$ in the $ab$ plane, the induced $\Omega_c^{(1)}$ reads

$$\Omega_c^{(1)}(n, \mathbf{k}) = (\partial_a G_{ba} - \partial_b G_{aa})E_a^{dc} + (\partial_a G_{bb} - \partial_b G_{ab})E_b^{dc}.$$

$D_\alpha^{(1)}$ defined in a few-layer 2D system can be approximately derived from $D_{\alpha c(\mathrm{bulk})}^{(1)}$ of

the bulk system by $D_\alpha^{(1)} = dD_{\alpha c(\text{bulk})}^{(1)}$, where $d$ is the thickness of the film. The independent components of $D_\alpha^{(1)}$ are related to the Berry connection polarizability tensor, $\mathbf{E}^{dc}$ and $d$. The mirror symmetry $M_a$ and the glide symmetry $\widetilde{M}_b$ in WTe$_2$ constrain $D_\alpha^{(1)}$ to be

$$D_a^{(1)} = \int_k [dk]\, f_0 [\partial_a(\partial_a G_{bb}) - \partial_a(\partial_b G_{ab})] \mathrm{E}_b^{dc} d,$$

$$D_b^{(1)} = \int_k [dk]\, f_0 [\partial_b(\partial_a G_{ba}) - \partial_b(\partial_b G_{aa})] \mathrm{E}_a^{dc} d,$$

where the other terms are prohibited by symmetry. In the experiment, the d.c. electric field is applied along a direction with an angle $\theta$ between $b$ axis, which can be expressed as $\mathbf{E}^{dc} = E^{dc}(-\sin\theta, \cos\theta)$. The induced BCD $\mathbf{D}^{(1)}(\theta) = \left(D_a^{(1)}(\theta), D_b^{(1)}(\theta)\right)$ hence reads

$$D_a^{(1)}(\theta) = \int_k [dk]\, f_0 [\partial_a(\partial_a G_{bb}) - \partial_a(\partial_b G_{ab})] \mathrm{E}^{dc} \cos\theta\, d,$$

$$D_b^{(1)}(\theta) = \int_k [dk]\, f_0 [\partial_b(\partial_b G_{aa}) - \partial_b(\partial_a G_{ba})] \mathrm{E}^{dc} \sin\theta\, d.$$

With the field-induced BCD, the second-order Hall current of an a.c. electric field $\mathbf{E}^\omega$ is [9]

$$j_\alpha^{2\omega} = -\varepsilon_{\alpha\mu\gamma} \frac{e^3\tau}{2(1+i\omega\tau)\hbar^2} D_{\beta\mu}^{(1)} E_\beta^\omega E_\gamma^\omega.$$

In two-dimensional systems, where $\mathbf{\Omega}^{(1)}$ is along out of plane direction and $D_{\alpha c}^{(1)} = \int_k [d\mathbf{k}] f_0 \left(\partial_\alpha \Omega_c^{(1)}\right)$, it is equivalent to

$$\mathbf{j}^{2\omega} = -\frac{e^3\tau}{2(1+i\omega\tau)\hbar^2} (\hat{\mathbf{z}} \times \mathbf{E}^\omega)[\mathbf{D}^{(1)}(\theta) \cdot \mathbf{E}^\omega].$$

The magnitude of induced second-order Hall conductivity is determined by $\mathbf{D}^{(1)}(\theta) \cdot \hat{\mathbf{E}}^\omega$, which is the projection of the pseudo vector $\mathbf{D}^{(1)}$ to the direction of $\mathbf{E}^\omega$, and the direction of Hall current is perpendicular to $\mathbf{E}^{\omega}$. Consequently, we can measure the $\mathbf{E}^{dc}$ induced BCD $\mathbf{D}^{(1)}$ by detecting its projective component $D_a^{(1)}(\theta)$ or $D_b^{(1)}(\theta)$ with an a.c. electric field along the corresponding direction. From the above derivation, when the direction of the d.c electric field varies in the $ab$ plane, the independent components of induced BCD $D_a^{(1)}$ and $D_b^{(1)}$ change as a cosine and a sine function, respectively. This relation is clearly demonstrated by our experimental results in Fig. 4 of main text.

With first-principles calculations, we estimate the extreme value of $D_a^{(1)}(0°)$ and $D_b^{(1)}(90°)$, as shown in **Fig. S10(d)**. It is taken that $d \sim 8.4$ nm and $E^{dc} \sim 3$ kV/m according to the experiment. $D_a^{(1)}(0°)$ and $D_b^{(1)}(90°)$ refer to $D_a^{(1)}$ and $D_b^{(1)}$ as the applied $E^{dc}$ along the $b$ axis and $-a$ axis, respectively. It is found that $D_b^{(1)}(90°)$ varies from ~-0.14 nm to 0 as tuning chemical potential away from 0, and $D_a^{(1)}(0°)$ shows a non-monotonic change between 0.18 and -0.13 nm as changing chemical potential. The experimental results of $D_b^{(1)}(90°)$ ~-0.05 nm and $D_a^{(1)}(0°)$ ~-0.28 nm (Fig. 4 in main text) agree well with the calculations on the order of magnitude.

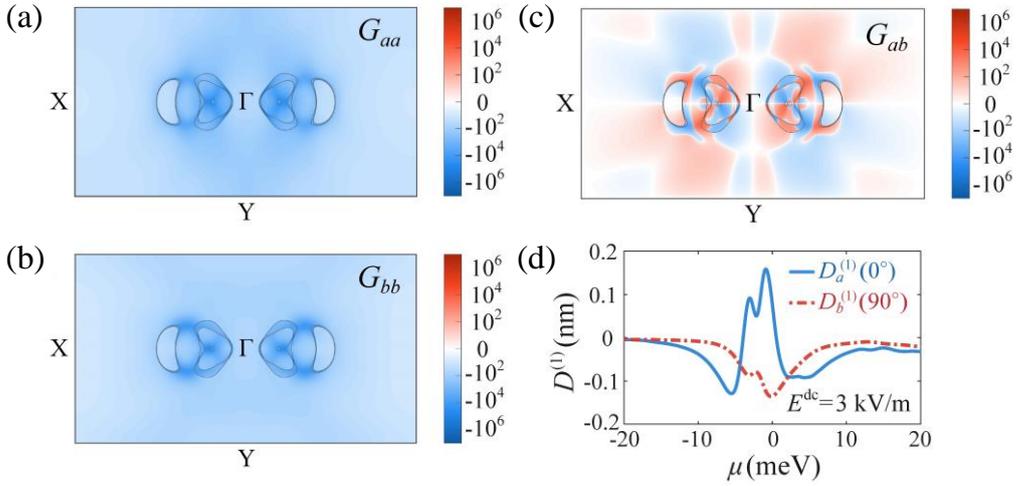

**Figure S10: Calculations of Berry connection polarizability tensor and field-**

induced Berry curvature dipole in WTe$_2$.

**a-c,** The calculated distribution of Berry connection polarizability tensor elements (**a**) $G_{aa}$, (**b**) $G_{bb}$, (**c**) $G_{ab}$ in the $k_z = 0$ plane of the Brillouin Zone for the occupied bands. The unit of BCP is $\text{Å}^2 \cdot \text{V}^{-1}$. The grey lines depict the Fermi surface.

**d**, Calculated field-induced BCD $D_a^{(1)}(0°)$ and $D_b^{(1)}(90°)$ with respect to the chemical potential $\mu$ when $E^{\text{dc}} = 3$ kV/m. In the calculations, the finite temperature effect is considered with a boarding of $k_B T$ at 5 K.

## Supplemental Note 7: Electric field dependence of second-order Hall signals.

The second-harmonic I-V characteristics in Fig. 1(e) of main text are converted into the $V_H^{2\omega}$ versus $(V^\omega)^2$ in **Fig. S11(a)**, where linear relationships are observed. The $\frac{E_H^{2\omega}}{(E^\omega)^2}$ as a function of the applied $E^{dc}$ is further calculated and presented in **Fig. S11(b)**.

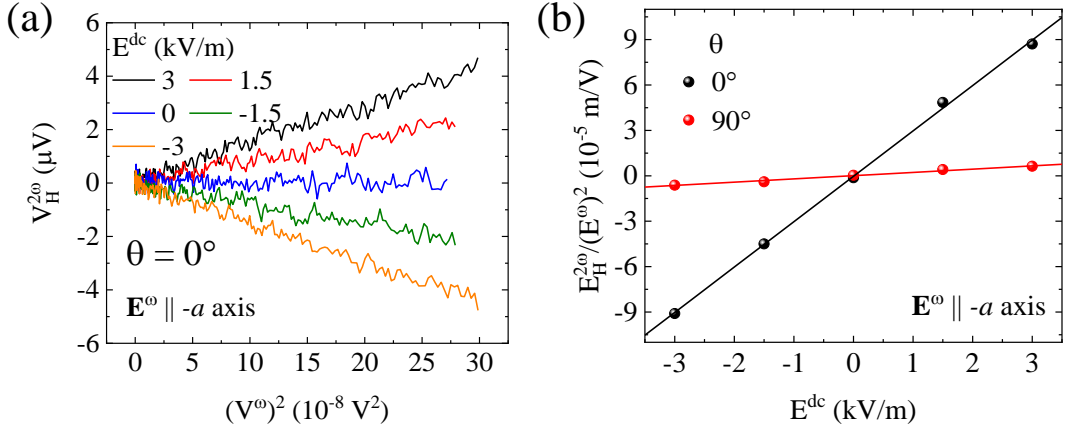

**Figure S11: Second-order AHE modulated by d.c. electric field at 5 K.**

**a**, The second-harmonic Hall voltage $V_H^{2\omega}$ as a function of $(V^\omega)^2$ as $\mathbf{E}^{dc}$ along $b$ axis and $\mathbf{E}^\omega$ along -$a$ axis.

**b**, The second-order Hall signal $\frac{E_H^{2\omega}}{(E^\omega)^2}$ as a function of $E^{dc}$ at $\theta = 0°$ and $\theta = 90°$ with $\mathbf{E}^\omega \parallel -a$ axis.

## Supplemental Note 8: Control experiments in device S2.

To demonstrate the symmetry constraint in WTe$_2$, control experiments were carried out in device S2. As schematically shown in **Figs. S12(a), (d)**, the a.c. and d.c. current sources are applied. The SR830 is an effective a.c. current source as connecting a resistor in series with output impedance 10 kΩ. The d.c. source is the Keithley current source with output impedance ~20 MΩ. For the d.c. field applied along $a$ and $b$ axis, respectively, the first-harmonic Hall voltage shows no obvious dependence on $\mathbf{E}^{dc}$, as shown in **Figs. S12(b)** and **S12(e)**, which indicate the independence of the two electric sources. When applying $\mathbf{E}^{dc} \parallel \mathbf{E}^{\omega} \parallel a$ axis, no second-order nonlinear Hall effect can be observed in **Fig. S12(c)**. Nevertheless, upon applying $\mathbf{E}^{dc} \perp \mathbf{E}^{\omega}$ and $\mathbf{E}^{\omega} \parallel a$ axis, as shown in **Fig. S12(f)**, nonzero second-order nonlinear Hall effect emerges due to the $\mathbf{E}^{dc}$ induced Berry curvature dipole along $a$ axis.

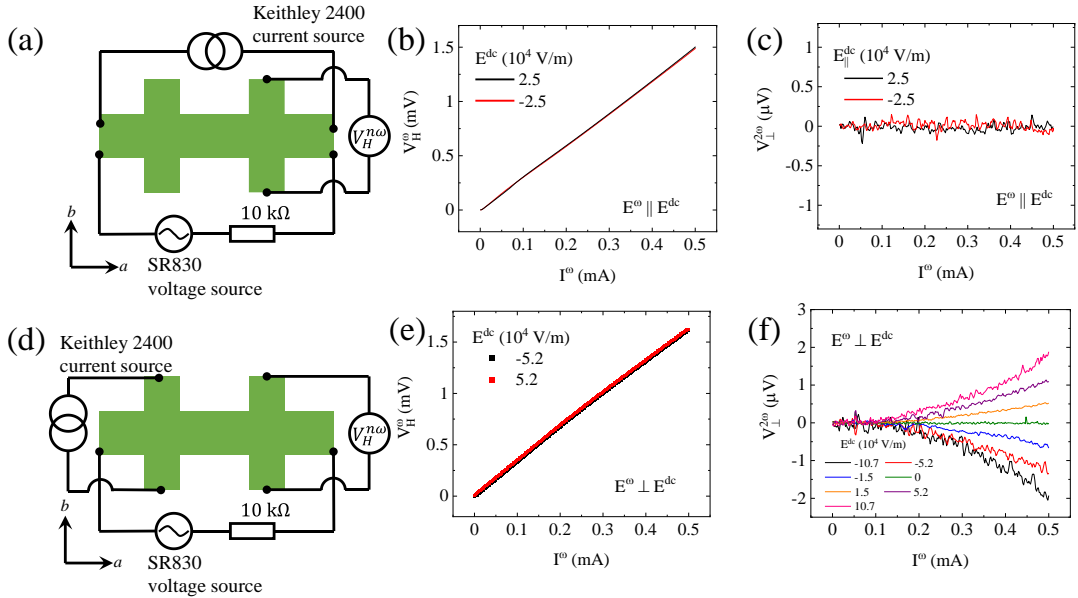

**Figure S12: The measurements by applying both d.c. electric field $\mathbf{E}^{dc}$ and a.c. current in devices S2 at 1.8 K.**

**a,** Schematic of the measurement configuration for (**b**) and (**c**).

**b,** First-harmonic Hall voltage $V_H^\omega$ under $\mathbf{E}^{dc} \parallel \mathbf{E}^\omega \parallel a$ axis.

**c,** There is no clear second-harmonic Hall voltage $V_H^{2\omega}$ under $\mathbf{E}^{dc} \parallel \mathbf{E}^\omega \parallel a$ axis.

**d,** Schematic of the measurement configuration for (**e**) and (**f**).

**e,** The $V_H^\omega$ under various $\mathbf{E}^{dc}$ with $\mathbf{E}^{dc} \perp \mathbf{E}^\omega$ and $\mathbf{E}^\omega \parallel a$ axis.

**f,** The $V_H^{2\omega}$ under various $\mathbf{E}^{dc}$ with $\mathbf{E}^{dc} \perp \mathbf{E}^\omega$ and $\mathbf{E}^\omega \parallel a$ axis.

**Supplemental Note 9: Discussions of other possible origins of the second order AHE.**

**1) Diode effect.** An accidental diode due to the contact can lead to a rectification, causing high-order transport, which, however, can be safely ruled out in this work due to the following reasons:

(a) Extrinsic signals of this origin should be strongly contact dependent. Thus, the angle-dependence should be also coupled to extrinsic contacts. Nevertheless, the angle-dependence of second-order AHE in Fig. 2 and **Fig. S12** is well consistent with the inherent symmetry of $WTe_2$, which excludes the extrinsic origins.

(b) The two-terminal d.c. measurements for all the diagonal electrodes show linear I-V characteristics, as shown in **Fig. S13(a)**, excluding the existence of diode effect. Linear fittings are performed for the two-terminal I-V curves. The R-square of the linear fittings is at least larger than 0.99997, indicating perfect linearity. Further, the deviation from linearity is analyzed by subtracting the linear-dependent part, as shown in **Fig. S13(b)**. It is found $\Delta V^{dc}$, *i.e.*, the deviation part, is four orders of magnitude smaller than the original $V^{dc}$, indicating a negligible nonlinearity. Moreover, the $\Delta V^{dc}$ shows no obvious current or angle dependence **(Fig. S13(b))**, and its magnitude is also much smaller than that of the higher-harmonic Hall voltages (**Fig. S13(c)**), further indicating that the observed higher-order transport in this work is failed to be attributed to the diode effect induced by contact.

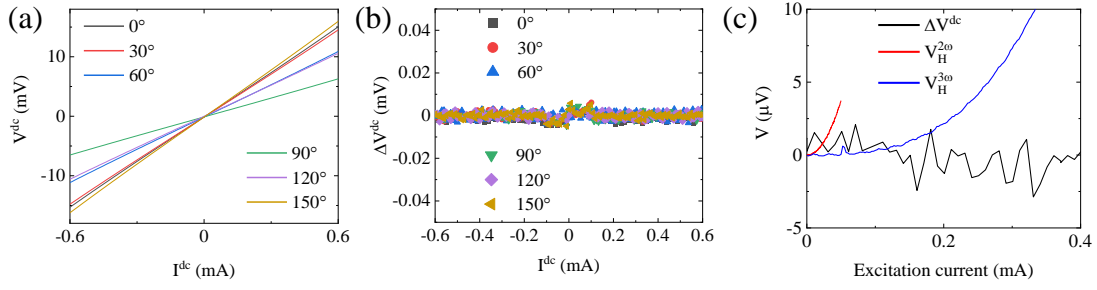

**Figure S13: Two-terminal d.c. measurements at 5 K in device S1.**

**a,** Current-voltage curves from two-terminal d.c. measurements for all the diagonal electrodes.

**b,** The current dependence of $\Delta V^{dc}$, that is, the deviations from the linearity of the current-voltage curves in Fig. S13a.

**c,** The comparation of the $\Delta V^{dc}$, $V_H^{2\omega}$ and $V_H^{3\omega}$. For $\Delta V^{dc}$ and $V_H^{3\omega}$, the excitation current is applied at $\theta = 30°$, while for $V_H^{2\omega}$, the excitation current is applied along $a$ axis and a d.c. field 3 kV/m is applied at $\theta = 30°$.

**2) Capacitive effect.** Contact resistance is generally inevitable between the metal electrodes and two-dimensional materials, which would induce an accidental capacitive effect, resulting in higher-order transport effect. Here, the second-order AHE shows a negligible dependence on frequency, as shown in **Fig. S14(a)**, excluding the capacitive effect. The phase of the second-harmonic Hall voltage is also investigated, where the Y signal dominates over the X signal (**Fig. S14(b)**). The phase of the second-harmonic Hall voltage is approximately $\pm 90°$, as shown in **Fig. S14(c)**. These features further exclude the capacitive effect.

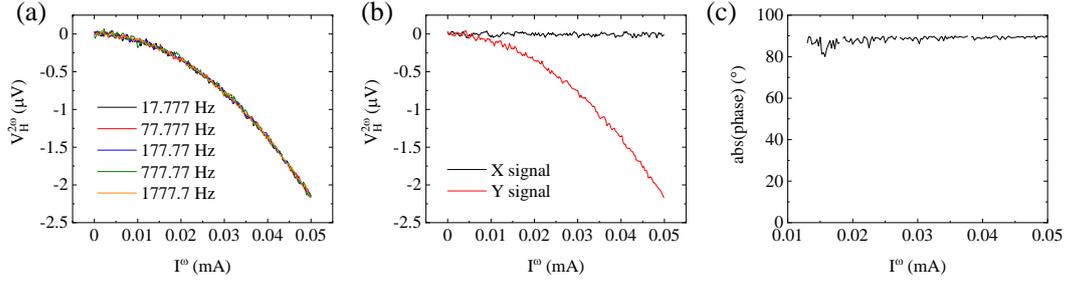

**Figure S14: Frequency-dependence and phase of second-order AHE in device S1 at 5 K and with $E^{dc} = 3$ kV/m at $\theta = 60°$.**

**a**, The second-order Hall signals at different frequencies.

**b**, The X and Y signals of the second-order Hall voltages.

**c**, The absolute value of the phase of the second-order Hall voltages.

**3) Thermal effect.** The thermal effect can also induce a second-order signal [41]. If the observed nonlinear Hall effect origins from thermal effect, it should response to both longitudinal and transverse d.c. electric field. However, as shown in **Fig. S12**, when applying $E^{dc} \parallel E^{\omega} \parallel a$ axis, no second-order nonlinear Hall effect is observed. Nevertheless, upon applying $E^{\omega} \perp E^{dc}$, nonzero second-order nonlinear Hall effect emerges. This observation is clearly inconsistent with the thermal effect. Moreover, the observed second-order nonlinear Hall effect shows strong anisotropy, as shown in Fig. 2 of main text. The angle-dependence of the d.c. field-induced second-order Hall effect is well consistent with the inherent symmetry of WTe$_2$, which is failed to be explained by the thermal effect.

**4) Thermoelectric effect.** Joule heating induced temperature gradient across the sample can drive a thermoelectric voltage, leading to second-order nonlinear Hall effect. This thermoelectric effect can also be excluded due to the following reasons:

(a) Uniform Joule heating will not induce a temperature gradient and thus no thermoelectric voltage across the sample.

(b) To generate thermoelectric voltage, the Joule heating should couple with external asymmetry, such as contact junction or flake shape, which should be unrelated to the inherent symmetry of WTe$_2$. However, the anisotropy of second-order nonlinear Hall effect is well consistent with the inherent symmetry analysis, as shown in Fig. 2 of main text.

5) **A residue of the first-harmonic Hall response $V_H^\omega$.** The influence of $V_H^\omega$ on the $V_H^{2\omega}$ can be ruled out because the first- and second-harmonic signals show different dependence on the d.c. electric field. As shown in **Fig. S15**, the first-harmonic Hall signal ($V_H^\omega$) shows that the I-V curves under $E^{dc} = \pm 3$ kV/m overlap with each other. By comparison, the second-harmonic Hall signal ($V_H^{2\omega}$) shows an anti-symmetric dependence on $E^{dc}$, where the sign of $V_H^{2\omega}$ is changed upon changing the sign of $E^{dc}$. This indicates that the existence of the first order signal $V_H^\omega$ will not affect the measurements of the second order signal $V_H^{2\omega}$.

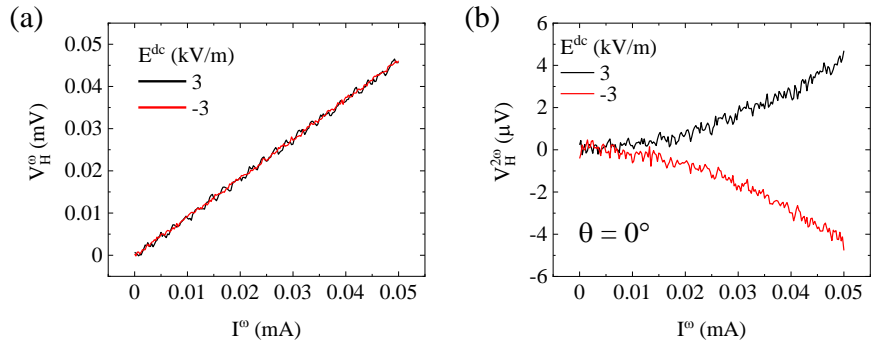

**Figure S15: The first- and second-harmonic signals at 5 K as $E^{dc}$ along $b$ axis ($\theta = 0°$) and $E^\omega$ along -$a$ axis.**

**a,** The first-harmonic Hall voltage $V_H^\omega$ as a function of $I^\omega$ at $E^{dc} = \pm 3$ kV/m.

**b,** The second-harmonic Hall voltage $V_H^{2\omega}$.

**6) Trivial effect by d.c. source.** We measured the first-harmonic longitudinal voltage upon applying $E^{dc} = 3$ kV/m, as shown in **Fig. S16**. It is clearly found that when reversing the sign of d.c. electric field, the I-V curves overlapped with each other. The results show that the d.c. source will not affect the a.c. measurements.

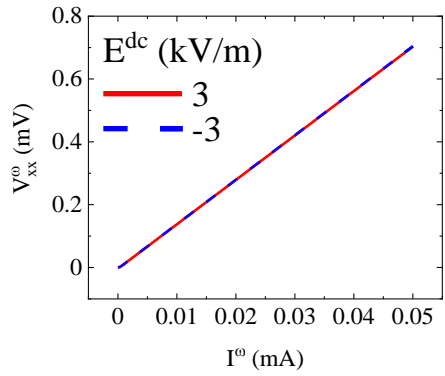

**Figure S16: The first-harmonic longitudinal voltage versus current under different d.c. electric fields at 5 K. The $E^{\omega}$ and $E^{dc}$ are along *a* axis.**

**7) Longitudinal nonlinearity originating from a circuit artifact.** We have measured both the second-harmonic Hall and longitudinal voltage at all the angles, as shown in **Fig. S17**. The measurement configuration is shown in the inset of **Fig. S17(d)** with d.c. field applied at angle $\theta$. It is clearly found that the Hall nonlinearity is dominated over longitudinal one, which guarantees that the observed second-order Hall effect doesn't originate from the longitudinal nonlinearity induced by a circuit artifact.

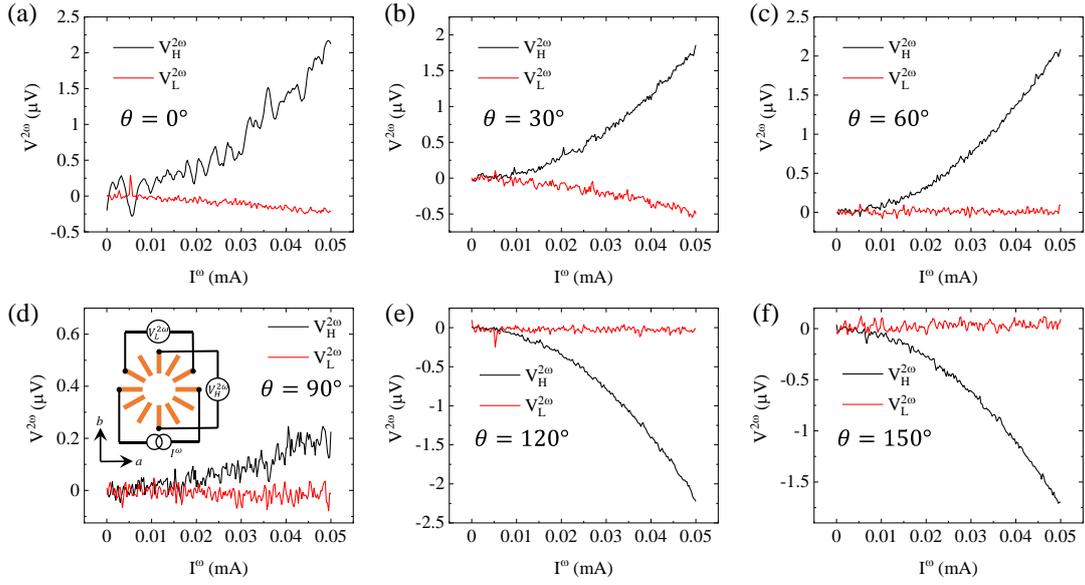

**Figure S17:** The second-harmonic Hall $V_H^{2\omega}$ and longitudinal voltage $V_L^{2\omega}$ with $E^\omega \parallel -a$ axis and $E^{dc} = 1.5$ kV/m along different angles at 5 K. The angle $\theta$ is defined in Fig. 1(d) of main text.

## Supplemental Note 10: Angle dependence of parameter $C_0$ obtained from the fittings of scaling law.

The second-order Hall signal $\frac{E_H^{2\omega}}{(E^\omega)^2}$ is found to satisfy scaling law $\frac{E_H^{2\omega}}{(E^\omega)^2} = C_0 + C_1 \sigma_{xx} + C_2 \sigma_{xx}^2$. For $\mathbf{E}^{dc} = 3$ kV/m with a fixed direction (angle $\theta$), a set of curves of $V_H^{2\omega}$ vs. $I^\omega$ is measured at different temperatures as $I^\omega$ is applied along -$a$ axis and $b$ axis, respectively. Through varying temperature, the $\sigma_{xx}$ is changed accordingly. Therefore, for a fixed angle $\theta$, the relationship between $\frac{E_H^{2\omega}}{(E^\omega)^2}$ and $\sigma_{xx}$ is plotted. By fitting the experimental data, the parameter $C_0$ is then obtained and presented in **Fig. S18**.

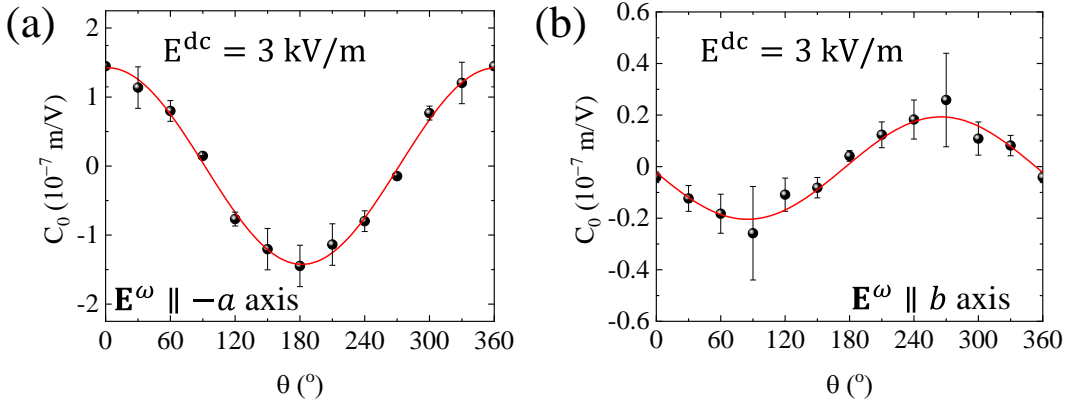

**Figure S18: Angle-dependence of the coefficient $C_0$.**

**a,b,** The coefficient $C_0$ as a function of $\theta$ with the amplitude of $\mathbf{E}^{dc}$ fixed at 3 kV/m for (**a**) $\mathbf{E}^\omega \parallel -a$ axis and (**b**) $\mathbf{E}^\omega \parallel b$ axis.